\documentclass[conference,a4paper]{IEEEtran}
 \pdfoutput=1 
\IEEEoverridecommandlockouts
\usepackage{cite}
\usepackage{amsmath,amssymb,amsfonts}
\usepackage{algorithmic}
\usepackage{graphicx}
\usepackage{textcomp}
\usepackage{xcolor}
\def\BibTeX{{\rm B\kern-.05em{\sc i\kern-.025em b}\kern-.08em
    T\kern-.1667em\lower.7ex\hbox{E}\kern-.125emX}}
\usepackage{tikz}
\tikzstyle{pinstyle} = [pin edge={to-,thin,black}]
\usetikzlibrary{calc}
\newlength\fheight
\newlength\fwidth
\usepackage{pgfplots}
\usepackage{cite}
\usepackage{url}
\usepackage{textcomp}
\usepackage[absolute]{textpos}

\newcommand{\copyrightstatement}{
    \begin{textblock}{15}(0.5,0.3)    
         \noindent
         \centering
         \textblockcolour{white}
         \footnotesize
         \copyright 2019 IEEE. Personal use of this material is permitted. Permission from IEEE must be obtained for all other uses, in any current or future media, including reprinting/republishing this material for advertising or promotional purposes, creating new collective works, for resale or redistribution to servers or lists, or reuse of any copyrighted component of this work in other works
    \end{textblock}
}

\begin{document}

\copyrightstatement
\title{Extending Video Decoding Energy Models for 360$^\circ$ and HDR Video Formats in HEVC}

\author{\IEEEauthorblockN{Matthias Kr\"anzler, Christian Herglotz, and Andr\'e Kaup}
	\IEEEauthorblockA{Chair of Multimedia Communications and Signal Processing, \\
		Friedrich-Alexander University Erlangen-N\"urnberg (FAU)\\
		\{matthias.kraenzler, christian.herglotz, andre.kaup\}@fau.de\\}}

\maketitle

\begin{abstract}
Research has shown that decoder energy models are helpful tools for improving the energy efficiency in video playback applications. For example, an accurate feature-based bit stream model can reduce the energy consumption of the decoding process. However, until now only sequences of the SDR video format were investigated. Therefore, this paper shows that the decoding energy of HEVC-coded bit streams can be estimated precisely for different video formats and coding bit depths. Therefore, we compare a state-of-the-art model from the literature with a proposed model. We show that bit streams of the 360$^\circ$, HDR, and fisheye video format can be estimated with a mean estimation error lower than $3.88\%$ if the setups have the same coding bit depth. Furthermore, it is shown that on average, the energy demand for the decoding of bit streams with a bit depth of 10-bit is $55\%$ higher than with 8-bit. 
\end{abstract}


\IEEEpeerreviewmaketitle

\section{Introduction}
\label{sec:intro}

According to recent studies, the global mobile data traffic for video content will rise from 12 exabytes per month in 2018 to 61 exabytes per month in 2022~\cite{CSI2019}. Furthermore, the study states that one key inhibitor of augmented reality (AR) and virtual reality (VR) content is the short battery lifetime of mobile devices. Therefore, the relevance for efficient transmission of 360$^\circ$ video content will rise in importance. 

Furthermore, the development of the next-generation video standard Versatile Video Coding (VVC) has been initiated with the goal of reducing the bit rate by $50\%$ with equal visual quality in comparison to HEVC~\cite{Sullivan2017}. This video standard will not only have a better compression performance, but it will also include special modes for new video formats like high dynamic range (HDR) and 360$^\circ$ content. Therefore, the test category standard dynamic range (SDR) is extended by these two video formats~\cite{JVET-H1002}.

For conventional SDR sequences, researchers developed a model to accurately estimate the energy consumption of the decoding process~\cite{HerglotzSpringerReichenbachEtAl2018}. If the model can estimate the decoding energy with sufficient accuracy, it can be exploited to reduce the energy demand of the device. It was shown that the energy consumption of the decoding process in the state-of-the-art video standard High Efficiency Video Coding (HEVC) can be reduced by up to $15\%$ with an equal visual objective quality, at the cost of increasing bit rate by less than $5\%$~\cite{HerglotzHeindelKaup}.

The decoding energy can be modeled with various approaches. One example are bit stream feature-based models, where features can be counted using a special analyzer software~\cite{Herglotz2019}. It could be shown that the estimation error of feature-based models is lower than $8\%$ for different video standards like HEVC, H.264, H.263, and VP9~\cite{HerglotzWenDaiEtAl2016}.

In other works, the Rate-Distortion Optimization (RDO) of the encoder was modified in such a way that modes with a high energy demand will be chosen with a lower probability~\cite{CorreaCorreaPalominoEtAl2018}. The energy-aware choice of modes resulted in energy savings of up to $17.7\%$ with a similar increase in bit rate. In~\cite{MallikarachchiTalagalaH.EtAl2017}, the complexity of the decoder was modeled by processor counts and several decoder functions. By that, the decoding complexity is calculated and the RDO is substituted by a complexity optimization, which aims to reduce the complexity of the decoder and thereby the energy demand, too.

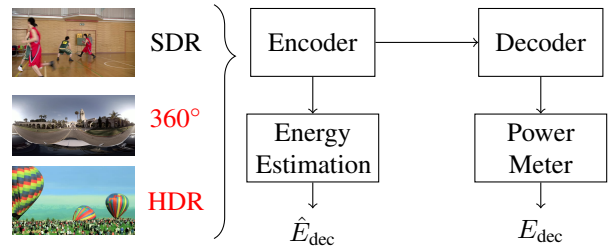
\begin{figure}[!t]
	\label{fig:BlockChart}
	\centering
	\setlength\fheight{4cm}
	\setlength\fwidth{7.5cm}
	\begin{tikzpicture}
	\begin{scope}[minimum width=15mm,minimum height=9mm,align=center]
	\node[draw,text width=1.4cm] (v1) at (0,0) {Encoder};
	\node[draw,text width=1.4cm] (v2) at ($(v1)+(3,0)$) {Decoder};
	\node[draw,text width=1.5cm] (v3) at ($(v1) + (0,-1.4)$) {Energy \\Estimation};
	\node[draw,text width=1.5cm] (v4) at ($(v2) + (0,-1.4)$) {Power \\Meter};
	\node[text width=1.5cm] (v5) at ($(v1) + (0,-2.5)$) {$\hat{E}_{\text{dec}}$};
	\node[text width=1.5cm] (v6) at ($(v2) + (0,-2.5)$) {$E_{\text{dec}}$};

	\node[text =black]  (v7) at (-1.8,0) {SDR};
	\node[text=red] (v8) at ($(v7) + (0,-1)$) {360$^\circ$};
	\node[text=red] (v9) at ($(v7) + (0,-2.1)$) {HDR};
	\node[inner sep=0pt] (image1) at ($(-2.75,1) + (-0.4,-1)$){\includegraphics[width=1.6cm]{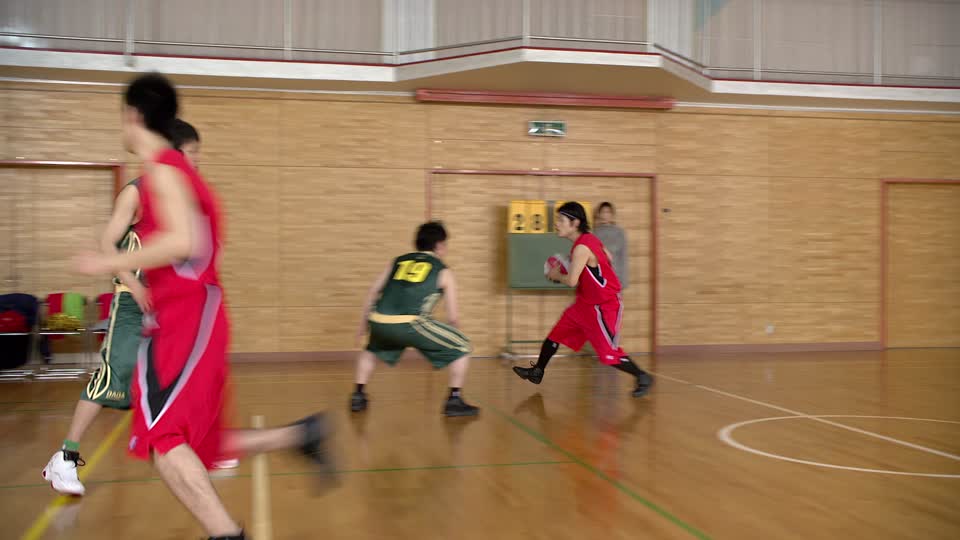}};
	\node[inner sep=0pt] (image2) at ($(image1) + (-0,-1.1)$){\includegraphics[width=1.6cm]{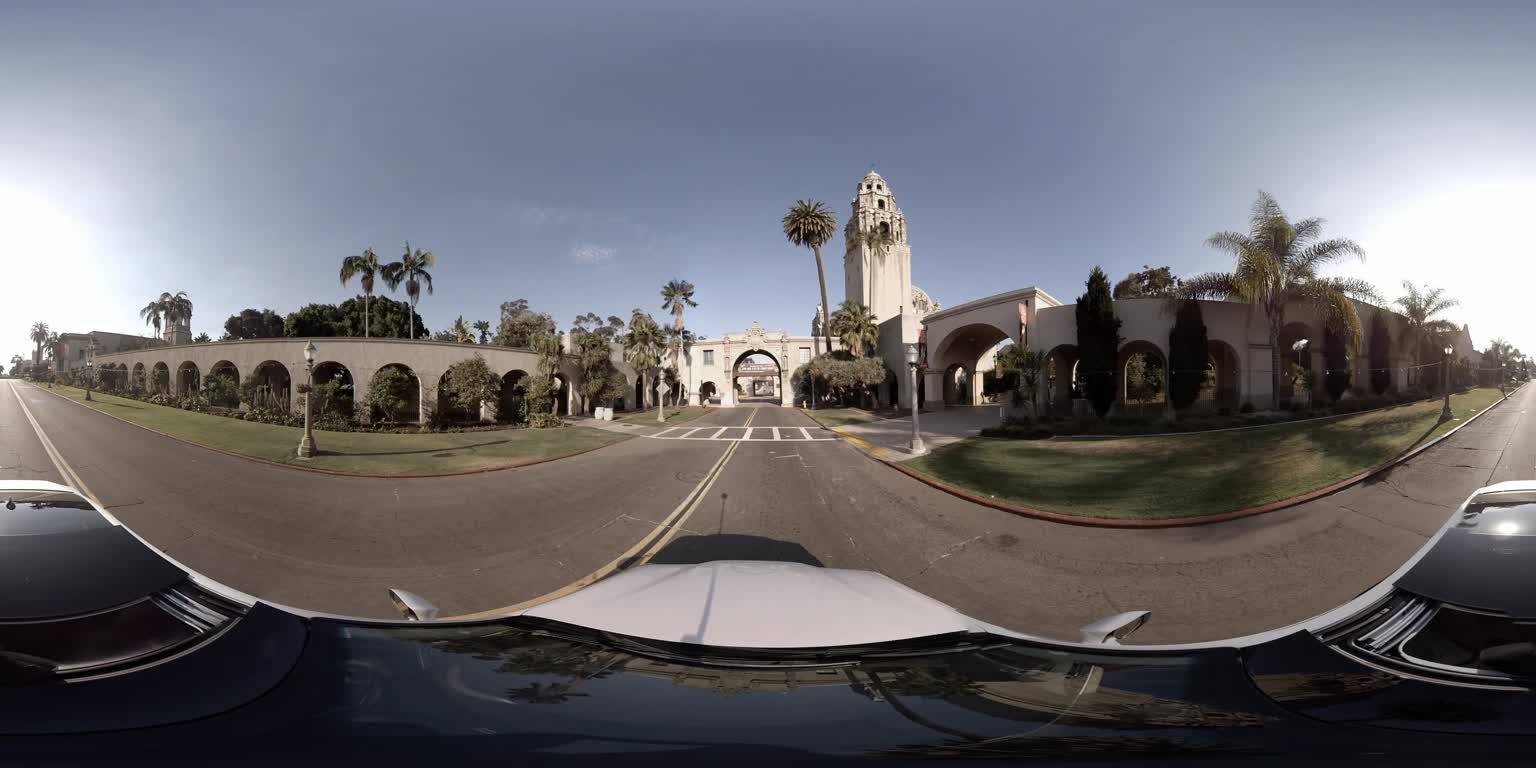}};
	\node[inner sep=0pt] (image3) at ($(image2) + (0,-1)$){\includegraphics[width=1.6cm]{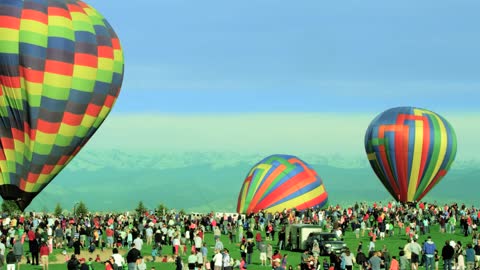}};

	\end{scope}

	\draw[->] (v1) -- (v2);
	\draw[->] (v1) -- (v3);
	\draw[->] (v3) -- ($(v5) + (0,0.3)$);
	\draw[->] (v2) -- (v4);
	\draw[->] (v4) -- ($(v6) + (0,0.3)$);
	
	\draw [decorate,decoration={brace,amplitude=8pt,aspect=0.22},yshift=0pt]
	($(v7) + (0.5,0.5) $) -- ($(v7) + (0.5,-2.6) $)  ;
	
	\end{tikzpicture}
	\vspace{-0.5cm}
	\caption{Overview of the complete evaluation toolchain with the newly evaluated video formats (360$^\circ$, HDR, and fisheye). First, the input sequences are encoded. Then, the decoding energy $E_{\text{dec}}$ is measured with the help of a power meter. Energy estimation returns the modeled energy $\hat{E}_{\text{dec}}$, which shall approximate the measured energy $E_{\text{dec}}$.
		}
	\vspace{-0.6cm}
\end{figure}

In this paper, the estimation accuracy for the video formats fisheye, HDR, and 360$^\circ$ will be investigated, as shown in Figure~\ref{fig:BlockChart}. Moreover, we will introduce new features to the bit stream feature-based model. As another contribution, our measurements show that the decoding energy significantly depends on the coding bit depth. Therefore, the energy model of the literature will be extended by two new parameters. With the combination of newly added features and the energy model with bit depth extension, we are able to estimate the decoding energy with sufficient accuracy regardless of the used video format or coding bit depth. Furthermore, the developed model reduces the estimation error significantly compared to the reference model.

The paper is organized as follows: First, Section~\ref{sec:basic} explains the fundamentals of feature-based energy modeling. In Section~\ref{sec:sequences}, the  setups for the different video formats are introduced. Afterwards, Section~\ref{sec:extended} presents the new features and the extension of the energy model for higher bit depth. In Section~\ref{sec:evaluation}, the used measurement setup will be presented and the results of the models and the setups will be discussed. Finally, Section~\ref{sec:conclusion} summarizes the finding of this paper.

\vspace{-0.1cm}
\section{Feature-Based Energy Model}
\label{sec:basic}

Previous research has found that the decoding energy can be accurately estimated by a bit stream feature-based energy model~\cite{HerglotzSpringerReichenbachEtAl2018}. In this context, a feature can be understood as a specific subprocess that is needed to decode a bit stream and finally reconstruct a frame. For example, the number of residuals, the prediction direction, the size of a transformation block, or the number of sample adaptive offset (SAO) blocks are potential subprocesses that influence the decoding energy. The estimated energy demand $\hat{E}_{\text{dec}}$, which a device needs to decode a certain bit stream, is defined as
\begin{equation}
\hat{E}_{\text{dec}} = \sum_{\forall i} n_{i} \cdot e_{i},
\label{eq:51}
\end{equation}
where $i$ indicates the index of a feature, the feature number $n_i$ counts how often the feature $i$ occurs within the bit stream, and $e_i$ is the specific energy coefficient of the feature $i$~\cite{HerglotzSpringerReichenbachEtAl2018}. The energy coefficient $e_i$ is related to the energy that is needed to process the feature during decoding. The derivation of each feature number $n_i$ can be executed both in the decoder and the encoder. An actual implementation of the decoder counting all features is available at~\cite{Herglotz2019}.

In this paper, we use the feature-based accurate (FA) model as a reference~\cite{HerglotzSpringerReichenbachEtAl2018} to compare with the proposed model, which will be introduced in Section~\ref{sec:extended}. Table~\ref{tab:FEModel} lists all features for all considered models, where the features from the FA model are marked by check marks. As can be seen from the table, all features are separated into five categories. Features that are mandatory for the decoding of a bit stream, such as the number of a certain frame type (e.g. Islice counts the number of intra frames) or the decoding energy that is needed for the initialization of the decoding process ($\text{E}_{\text{O}}$), are classified in the category `General'. The category `Intra' contains all features related to intra-frame prediction. Some features (e.g. pla) depend on the coding unit (CU) size of a block. Therefore, squared blocks sizes are categorized into depths with decreasing block size order from 0 to 4. In this context, depth 0 corresponds to a block size of 64$\times$64 and the depth 4 to a block size of 4$\times$4.

The category `Inter' comprises all features concerning different prediction modes (inter, skip, or merge), fractional pixel filters, and motion vectors. The features for residual coefficients and transformation are comprised in the category `Residual'. Finally, the category `In-loop' counts all executions of the deblocking filters (Bs0-2) and the SAO-filter. A detailed description of each feature can be found in~\cite{HerglotzSpringerReichenbachEtAl2018}.
\begin{table}[!t]
	\caption[FU model feature list.]{List of all features of the FU and the FA energy model. The features are labeled with corresponding depths and are assigned to categories. FA indicates features that are used in the FA model from~\cite{HerglotzSpringerReichenbachEtAl2018}. The features Bslice and Pslice add up to the feature PBslice in the FA model. $\phi_i$ indicates whether the specific energy $e_i$ of a feature is affected by the bit depth ($\phi_i$=1) or not ($\phi_i$=0).
	}
	\centering
	\bgroup
	\def\arraystretch{1.05}
	\vspace{-0.2cm}
	\begin{tabular}{l c c c c}
		\multicolumn{1}{l}{Feature Label} & Depths & FA & FU & ~~$\phi$~~\\
		\hline
		\multicolumn{5}{c}{General} \\
		\hline
		
		$\text{E}_{\text{O}}$  & - & $\checkmark$ & $\checkmark$ & $1$ \\
		Islice                 & - & $\checkmark$ & $\checkmark$ & $1$ \\
		Bslice, Pslice     & - & ($\checkmark$) & $\checkmark$ & $1$ \\
		\hline
		\multicolumn{5}{c}{Intra} \\
		\hline
		
		intraCU                & -     & $\checkmark$ & $\checkmark$ & $1$ \\
		pla, dc, hvd, ang     & 1 - 4 & $\checkmark$ & $\checkmark$ & $1$ \\
		noMPM                 & -     & $\checkmark$ & $\checkmark$ & $0$ \\
		\hline
		\multicolumn{5}{c}{Inter} \\
		\hline
		
		skip, merge, mergeSMP  & 0 - 3 & $\checkmark$ & $\checkmark$ & $1$ \\
		mergeAMP              & 0 - 2 & $\checkmark$ & $\checkmark$ & $1$ \\
		inter, interSMP       & 0 - 3 & $\checkmark$ & $\checkmark$  & $1$ \\
		interAMP              & 0 - 2 & $\checkmark$ & $\checkmark$ & $1$ \\
		fracpelHor, -Ver,     & 0 - 3 & $\checkmark$ & $\checkmark$  & $1$ \\
		fracpelBoth, copyPel      & 0 - 3 & - & $\checkmark$ & $1$ \\%
		chrHalfpel           & 0 - 3 & $\checkmark$ & $\checkmark$ & $1$ \\
		bi                  & -     & $\checkmark$ & $\checkmark$ & $1$ \\
		uni             & -     & - & $\checkmark$  & $1$ \\%
		MVD                 & -     & $\checkmark$ & $\checkmark$ & $0$ \\
		\hline
		\multicolumn{5}{c}{Residual} \\
		\hline
		
		coeff, coeffG1, val          & -     & $\checkmark$ & $\checkmark$ & $0$ \\
		CSBF                   & -     & $\checkmark$ & $\checkmark$ & $1$ \\
		TrIntraY/C, TrInterY/C & 1 - 4 & $\checkmark$ & $\checkmark$ & $1$ \\
		TSF                    & -     & $\checkmark$ & $\checkmark$ & $0$ \\
		\hline
		\multicolumn{5}{c}{In-loop} \\
		\hline
		Bs0, Bs1, Bs2              & -   & $\checkmark$ & $\checkmark$ & $1$ \\
		SAO Y BO/EO  & -   & $\checkmark$ & $\checkmark$ & $0$ \\
		SAO C BO/EO  & -   & $\checkmark$ & $\checkmark$  & $0$ \\
		SAO allComps  & -   & $\checkmark$ & $\checkmark$  & $0$ 
	\end{tabular}
	\label{tab:FEModel}
	\egroup
	\vspace{-0.7cm}
\end{table}
\vspace{-0.1cm}
\section{Sequence Setups}
\label{sec:sequences}

For the evaluation of the feature models, we use the common test conditions of various video formats. For the encoding of the bit streams, the HEVC-test model reference software (HM)~\cite{HM16} version 16.16 was used. For each original sequence, four different QPs (22, 27, 32, and 37) and configurations~(intra, lowdelay, lowdelayP, and randomaccess) are used. In Table~\ref{tab:Setups}, the number of original sequences, the number of bit streams, the used bit depth, and the source of the corresponding sequences are listed.

First, the Conventional8 and the Conventional10 setup represent the conventional rectilinear video format with a standard dynamic range (SDR). Both setups use the sequences of the HEVC test set from Class A-F~\cite{JCTVC-L1100} and \mbox{Class~A1-A2~\cite{JVET-K1010}}. Furthermore, the sequences of the Conventional8 setup are coded with an internal bit depth of 8-bit and for the Conventional10 setup with 10-bit, respectively. Each setup has 480 different bit streams. Due to the high resolution, the sequences of the classes A1 and A2 are encoded with 25 frames and sequences of the class A with 50 frames. For the other classes, the recommended number of frames is used.

For HDR, we use the recommendations of~\cite{JVET-K1011}, which suggest 11 sequences (see Table~\ref{tab:Setups}). Therefore, each setup has 176 bit streams. Depending on the coding bit depth, the setup with 8-bit is called HDR8, and the setup with 10-bit is called HDR10, respectively. Due to the 4K resolution, sequences of class H2 are only encoded with 25 frames.

For fisheye video content, a data set with 62 unique sequences can be found in~\cite{Eichenseer2016}. These sequences are separated into synthetic and real-world sequences. The former sequences are cropped to a resolution of 1072$\times$1072 and the latter sequences to a resolution of 1136$\times$1072. In total, the setup has 992 bit streams and, as we only considered a bit depth of 8-bit for encoding, the setup is called Fisheye.

The recommendations for 360$^\circ$ content are based on~\cite{JVET-K1012}. The 360$^\circ$ source sequences are given in the equirectangular projection~(ERP) format with an 8K (Class S1) or 6K (Class~S2) resolution and are shown in Table~\ref{tab:Setups}. For our measurements, these sequences are downsampled to a lower resolution and converted to different projection formats with the 360Lib software~\cite{360Lib}. In this paper, we considered six different projection formats of~\cite{JVET-K1012}. The downsampled sequences in the ERP format, padded equirectangular projection~(PERP), cubemap projection~(CMP), equi-angular cubemap projection (EAC), adjusted cubemap projection (ACP), and rotated sphere projection (RSP) format. A detailed description of each projection format can be found in the documentation of the 360Lib~\cite{360Lib}. The bit streams were downsampled to a 4K resolution, where for each projection format we chose the recommended resolution from~\cite{JVET-K1012}. The bit streams for the 360$^\circ$ video format are both coded at a bit depth of 8-bit for the 360D8 setup or with 10-bit for the 360D10 setup. Both setups have 960 bit streams and every sequence is coded with 25 frames.

\begin{table}[!t]
	\caption{List of all setups with the corresponding number of source sequences, the number of bit streams, the used bit depth, and the sources of the sequences.}
	\centering
	\bgroup
	\vspace{-0.2cm}
	\def\arraystretch{1}
	\begin{tabular}{c | c | c | c | l} 
		Setup          & \#Sequences & \#Bit Streams & Bit Depth & Source \\
		\hline           
		\hline     
		Conventional8  & 30 & 480 & 8  & \cite{JCTVC-L1100}, \cite{JVET-K1010} \\	
		Conventional10 & 30 & 480 & 10 & \cite{JCTVC-L1100}, \cite{JVET-K1010} \\
		Fisheye        & 62 & 992 & 8  & \cite{Eichenseer2016} \\
		360D8          & 10 & 960 & 8  & \cite{JVET-K1011} \\
		360D10         & 10 & 960 & 10 & \cite{JVET-K1011} \\
		HDR8           & 11 & 176 & 8  & \cite{JVET-K1012} \\
		HDR10          & 11 & 176 & 10 & \cite{JVET-K1012} \\
	\end{tabular}
	\label{tab:Setups}
	\egroup
	\vspace{-0.6cm}
\end{table}
\vspace{-0.1cm}
\section{Extended Decoder Energy Model}
\label{sec:extended}

This section introduces the proposed additional features in Section~\ref{subsec:FU}. Furthermore, Section~\ref{subsec:3B} presents the proposed extension for high bit depth coding.

\subsection{Feature-Based Universal Model}
\label{subsec:FU}
In the following, we introduce the feature-based universal model (FU) with additional features. In Table~\ref{tab:FEModel}, all features of the FU model are listed. First, the features Bslice and Pslice substitute the feature PBslice of the FA model, which is the sum of both features. Therefore, the number of  B-frames is counted for the feature Bslice and the number of P-frames for the feature Pslice.

In the FA model, the features fracpelHor and fracpelVer correspond to the luma pels of the prediction unit (PU) that have to be filtered if at least one dimension of a motion vector points towards a fractional pel position~\cite{HerglotzSpringerReichenbachEtAl2018}. To explain the features fracpelHor and fracpelVer as well as the new features fracpelBoth and copyPel, we distinguish for cases. In the first case, the motion vector has an integer length in horizontal dimension and a fractional length in vertical dimension. For this case, the feature number $n_{\text{fracpelVer}}$ is incremented by the number of luma pels in the current PU. Likewise, $n_{\text{fracpelHor}}$ is incremented if only the horizontal dimension points to a fractional position. For the third case, the motion vector has a fractional length in both horizontal and vertical dimension. In this case, the luma pels of the PU are added for $n_{\text{fracpelHor}}$ and for $n_{\text{fracpelVer}}$. Furthermore, in HEVC, six extra rows of filtering operations are performed outside of the borders of the PU~\cite{HerglotzSpringerReichenbachEtAl2018}. For the FA model, the feature $n_{\text{fracpelHor}}$ is incremented by $6 \cdot w$, where $w$ is the width of the PU. For the FU model, the additional pels are assigned to $n_{\text{fracpelBoth}}$. For the last case, the motion vector has an integer length in both dimensions. Here, we increment $n_{\text{copyPel}}$ by the number of luma pels within the PU.

Finally, uni is the last introduced feature of the FU model. In HEVC, a block can be predicted from one or two reference frames (uni- and bi-prediction, respectively)~\cite{Sze2014}. Therefore, $n_{\text{uni}}$ counts every 4$\times$4-subblock that is uni-predicted and, consequently, is the counterpart to the feature bi.

\subsection{Extension for Higher Bit Depth}
\label{subsec:3B}
\begin{figure}[!t]
	\centering
	\setlength\fheight{3.3cm}
	\setlength\fwidth{7.5cm}
	\input{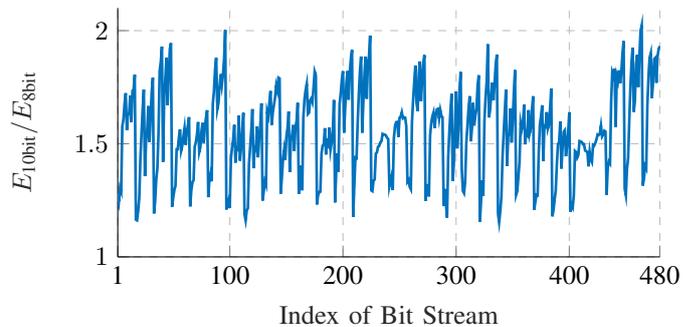}
	\vspace{-0.8cm}
	\caption{Comparison of the measured decoding energies $E_{\text{8bit}}$ and $E_{\text{10bit}}$  of bit streams with a coding bit depth of 8-bit (Conventional8 setup) and 10-bit (Conventional10 setup), respectively.}
	\label{fig:EnergyComp}
	\vspace{-0.6cm}
\end{figure}

By measurements, we found that the coding bit depth has significant influence on the decoding energy. To show this, we use the Conventional8 and Conventional10 setup. Both setups use the same input sequences and encoder configurations, but use different internal bit depths. In Figure~\ref{fig:EnergyComp}, the measured decoding energy of bit streams with a coding bit depth of 10-bit ($E_{\text{10bit}}$) is directly compared to the corresponding bit streams coded with 8-bit ($E_{\text{8bit}}$). The graph shows that bit streams with 10-bit have a significantly higher decoding energy, some bit streams even require more than twice the decoding energy of the corresponding bit stream with 8-bit. However, the ratio has a high variance with a difference between $15\%$ and $102\%$. Averaging over the given test set, 10-bit decoding consumes $55\%$ more energy than 8-bit decoding.

This difference in decoding energy leads to problems in decoding energy modeling, because the dependency between bit depth and specific energy coefficient $e_i$ is not considered in \eqref{eq:51}. Therefore, we extend the energy model by
\begin{equation}
\hat{E}_{\text{dec,10bit}} = \sum_{\forall i} \left( 1 + \zeta \cdot \phi_{i}  \right) \cdot e_{\text{8bit,}i} \cdot n_{i},
\label{eq:isInfluenced}
\end{equation}
where $e_{8bit,i}$ is the specific energy coefficient for a bit depth of \mbox{8-bit}, $\phi_{i}\in\{0,1\}$ a parameter that indicates whether the energy coefficient of a feature $i$ depends on the coding bit depth, and $\zeta$ represents a scaling factor. 

For the determination of $\phi_i$, we used a brute-force algorithm to find the best solution. We separated all 100 features from FU into groups of similar features, e.g. all features concerning the intra-prediction direction (pla, dc, hvd, and ang), in order to reduce the computing complexity. The brute-force was trained with the Conventional8 setup and validated with the Conventional10 setup. 

\vspace{-0.1cm}
\section{Evaluation}
\label{sec:evaluation}
\vspace{-0.1cm}
\subsection{Measurement Setup}
The setup to measure the decoding energy of the bit streams is based on~\cite{HerglotzSpringerReichenbachEtAl2018} and consists of three components. The power demand is measured with a high precision power meter (ZES Zimmer LMG 95), the device-under-test (DUT) is a Raspberry Pi 3B+~\cite{Raspberry}, and the voltage source is a HAMEG HM7042, which supplies the DUT with power. In order to obtain the decoding energy, the power meter measures the current through and the voltage across the main supply jack of the Raspberry Pi. The decoding energy is determined by two measurements. At first, the energy in idle mode is measured and then the energy during the decoding process of the bit stream. By subtraction, the directly caused energy of the decoding process is determined. To ensure the statistical correctness of the measurement, multiple measurements are performed for each bit stream and a confidence interval test described in~\cite{HerglotzSpringerReichenbachEtAl2018} is applied.

The Raspberry Pi 3B+ is a single-board computer with a Cortex-A53 quadcore CPU, which has a clock frequency of 1.4~GHz~\cite{Raspberry}. The device has a smartphone-like architecture with the ARMv8 processor. As an operating system (OS), we used Raspbian Stretch Lite~\cite{Raspbian2} with the Kernel version 4.14.71. The OS only has a terminal interface, because a graphical user interface (GUI) would cause a higher power consumption and more background processes, which have a disturbing influence on the measurement accuracy.

The decoding of the bit streams is performed using the FFmpeg framework~\cite{FFmpeg} version 4.0.3. FFmpeg is developed for practical real-time applications and is capable of using all cores of the Raspberry simultaneously~\cite{HerglotzHeindelKaup}.

\subsection{Training and Validation}
For the evaluation of the estimation, we calculate the mean estimation error $\bar{\varepsilon}$, which is defined by
\begin{equation}
\bar{\varepsilon} = \frac{1}{L} \sum_{l=1}^{L} \left| \frac{ \hat{E_{l}} - E_{l\text{,dec}}  }{E_{l\text{,dec}}} \right|,
\label{eq:mee}
\end{equation}
where $l$ is the index of the bit stream, $L$ the number of bit streams within the setup, $\hat{E_{l}}$ the estimated decoding energy of a bit stream, and $E_{l\text{,dec}}$ the corresponding measured decoding energy of the bit stream~\cite{HerglotzSpringerReichenbachEtAl2018}.

To train the coefficients $e_i$, we use a trust-region-reflective algorithm with least squares fitting~\cite{ColemanLi1996}. Therefore, the determined feature numbers $n_i$ for each bit stream and the corresponding measured decoding energies $E_{\text{dec}}$ are used as input data for the training and optimized in terms of $\bar{\varepsilon}$. Reported estimation errors for a certain validation setup are calculated with trained coefficients $e_i$ from a disjunct training set.
\vspace{-0.12cm}
\subsection{Results for Additional Features}
\begin{table}[!t]
	\caption{Mean estimation error $\bar{\varepsilon}$ for the FA and the FU model with the training of the Conventional8 and Conventional10 setup.}
	\centering
	\bgroup
	\vspace{-0.2cm}
	\def\arraystretch{1}
	\begin{tabular}{c | c | c | c | c}
						 & \multicolumn{4}{c}{ Training Setup} \\
	                     & \multicolumn{2}{c|}{Conventional8} & \multicolumn{2}{c}{Conventional10} \\ 
	Validation Setup & FA \cite{HerglotzSpringerReichenbachEtAl2018}             & FU        & FA \cite{HerglotzSpringerReichenbachEtAl2018}       & FU\\
		\hline   
		\hline
	Conventional8        &      -          &      -    & $59.46\%$ & $60.21\%$ \\
	Fisheye              &  $~6.48\%$      & $~3.88\%$ & $60.24\%$ & $62.91\%$ \\
	360D8                &  $~4.18\%$      & $~2.58\%$ & $60.29\%$ & $61.12\%$ \\
	HDR8                 &  $~4.34\%$      & $~3.86\%$ & $68.89\%$ & $69.17\%$ \\
	\hline
	Conventional10       &  $35.95\%$      & $35.81\%$ &   -       &  -        \\
	360D10               &  $38.29\%$      & $37.86\%$ & $~3.37\%$ & $~1.73\%$ \\
	HDR10                &  $40.38\%$      & $40.32\%$ & $~3.48\%$ & $~2.80\%$
	\end{tabular}
	\label{tab:FirstResults}
	\egroup
	\vspace{-0.6cm}
\end{table}
For evaluation, we strictly separate the training from the validation data. Therefore, we evaluate the estimation error for all setups after training with a conventional setup. Table~\ref{tab:FirstResults} shows the mean estimation error $\bar{\varepsilon}$ for the FA and the FU model, where the conventional setup is used for coefficient training. The results of the table show that the proposed model improves the estimation error significantly. At first, we take a look at the results with the trained energy coefficients of the Conventional8 setup. For the FA model, $\bar{\varepsilon}$ is $6.48\%$ for the Fisheye setup, $4.18\%$ for the 360D8 setup, and $4.34\%$ for the HDR8 setup. With the FU model, we reduce $\bar{\varepsilon}$ to $3.88\%$, which corresponds to an absolute improvement of the mean estimation error $\Delta\bar{\varepsilon}$ of $2.6\%$. Furthermore, the 360D8 setup and HDR8 have both an improved $\bar{\varepsilon}$ of $2.58\%$ and $3.86\%$, respectively.

However, as expected from the direct comparison of measurements in Figure~\ref{fig:EnergyComp}, the setups with a higher coding bit depth show a significantly higher estimation error. For all three setups, the $\bar{\varepsilon}$ is between $35\%$ and $40\%$ for both models. A closer inspection of the estimation reveals that the estimated decoding energy $\hat{E}_{l}$ of the 8-bit sequences is always too low.

Nevertheless, the setups with a bit depth of 10-bit can be accurately estimated with the training of the Conventional10 setup. For the 360D10 setup, $\bar{\varepsilon}$ is $3.37\%$ for the FA model and $1.73\%$ for the FU model. Furthermore, for the HDR10 setup, $\bar{\varepsilon}$ is $3.48\%$ with the FA model and $2.80\%$ with the FU model. Again, the FU model improves the estimation error significantly. Though, for the remaining setups, $\bar{\varepsilon}$ is approximately $60\%$, which is caused by the increased energy demand of bit streams with higher bit depth.

\vspace{-0.1cm}
\subsection{Results for 10-bit Extension}
\label{subsec:10bit}
With both conventional setups and the FU model, it is possible to estimate all setups with $\bar{\varepsilon}$ lower than $3.88\%$. However, the measurement of bit streams is rather time consuming. In order to reduce the amount of time to measure both setups, we only use the training of the Conventional8 setup in the following to model the 10-bit setups.

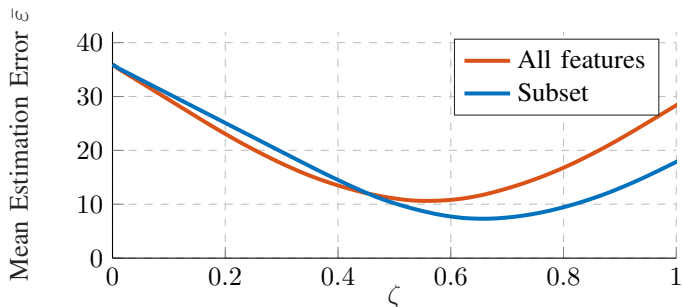
\begin{figure}[!t]
	\centering
	\setlength\fheight{3cm}
	\setlength\fwidth{7.8cm}
%
%
\definecolor{mycolor1}{rgb}{0.00000,0.44700,0.74100}%
\definecolor{mycolor2}{rgb}{0.85000,0.32500,0.09800}%
\begin{tikzpicture}

\begin{axis}[%
width=0.951\fwidth,
height=\fheight,
at={(0\fwidth,0\fheight)},
scale only axis,
xmin=0,
xmax=1,
xlabel style={font=\color{white!15!black}},
xlabel={$\zeta$},
xlabel style={yshift=0.3cm}, 
ymin=0,
ymax=42,
ylabel style={font=\color{white!15!black}},
ylabel={Mean Estimation Error $\bar{\varepsilon}$},
xmajorgrids,
xminorgrids,
ymajorgrids,
yminorgrids,
grid style={dashed},
axis x line*=bottom,
axis y line*=left,
axis background/.style={fill=white},
legend pos=north east,
legend style={legend cell align=left, align=left, draw=white!15!black}
]

\addplot [color=mycolor2, line width=1.5pt]
table[row sep=crcr]{%
	0    35.95\\
	0.01	35.1714193282379\\
	0.02	34.5295521928739\\
	0.03	33.8876850575099\\
	0.04	33.2458179221459\\
	0.05	32.6039507867819\\
	0.06	31.962083651418\\
	0.07	31.320216516054\\
	0.08	30.67834938069\\
	0.09	30.036482245326\\
	0.1	29.394615109962\\
	0.11	28.752747974598\\
	0.12	28.110880839234\\
	0.13	27.4690137038701\\
	0.14	26.8271465685061\\
	0.15	26.1852794331421\\
	0.16	25.5434122977782\\
	0.17	24.904833716158\\
	0.18	24.2764279820198\\
	0.19	23.6648507493086\\
	0.2	23.0610599016952\\
	0.21	22.4635906815985\\
	0.22	21.871244547532\\
	0.23	21.2914691089649\\
	0.24	20.7246555779483\\
	0.25	20.172211204924\\
	0.26	19.6319893954289\\
	0.27	19.0998628349732\\
	0.28	18.5781868502528\\
	0.29	18.0694869486575\\
	0.3	17.57288872506\\
	0.31	17.0887545489592\\
	0.32	16.6214808048634\\
	0.33	16.1693885871707\\
	0.34	15.7255107850635\\
	0.35	15.3060622443761\\
	0.36	14.9056546852828\\
	0.37	14.5296454652677\\
	0.38	14.1658190104506\\
	0.39	13.82258981367\\
	0.4	13.4930731148933\\
	0.41	13.1745461551239\\
	0.42	12.8749537460504\\
	0.43	12.5882096189382\\
	0.44	12.3175258476766\\
	0.45	12.0604204481684\\
	0.46	11.8257838472268\\
	0.47	11.6156281765343\\
	0.48	11.4214463044334\\
	0.49	11.2460960434647\\
	0.5	11.0837889985295\\
	0.51	10.9359201394485\\
	0.52	10.8084431348858\\
	0.53	10.7149081622979\\
	0.54	10.6553028976586\\
	0.55	10.6255742964857\\
	0.56	10.6167625880503\\
	0.57	10.6369831071368\\
	0.58	10.6745880901864\\
	0.59	10.7341694639996\\
	0.6	10.8095733925366\\
	0.61	10.9132110355871\\
	0.62	11.0431236435759\\
	0.63	11.1997239252694\\
	0.64	11.3811219862788\\
	0.65	11.5894218856377\\
	0.66	11.8149979636722\\
	0.67	12.0687649479503\\
	0.68	12.3400882698582\\
	0.69	12.6314452725803\\
	0.7	12.9358379015962\\
	0.71	13.2514202457217\\
	0.72	13.5781765298323\\
	0.73	13.9215098119681\\
	0.74	14.2854603299374\\
	0.75	14.6622681943496\\
	0.76	15.0518010273093\\
	0.77	15.4582088369286\\
	0.78	15.8834632148541\\
	0.79	16.3221565945954\\
	0.8	16.7748453161941\\
	0.81	17.2476653698415\\
	0.82	17.738429424119\\
	0.83	18.24308122722\\
	0.84	18.7603825800735\\
	0.85	19.2900874973483\\
	0.86	19.8286267636447\\
	0.87	20.3851507107282\\
	0.88	20.9556427085733\\
	0.89	21.53520892628\\
	0.9	22.1282754608157\\
	0.91	22.7321689579597\\
	0.92	23.3444376229367\\
	0.93	23.9643078570101\\
	0.94	24.5883531186342\\
	0.95	25.2140509396226\\
	0.96	25.843317717366\\
	0.97	26.4755955711105\\
	0.98	27.1091434335696\\
	0.99	27.7426912960287\\
	1	28.3799184758933\\
	1.01	29.0184150341626\\
	1.02	29.6582247275333\\
	1.03	30.2990284788883\\
	1.04	30.9408956142523\\
	1.05	31.5827627496162\\
	1.06	32.2246298849802\\
	1.07	32.8664970203442\\
	1.08	33.5083641557082\\
	1.09	34.1502312910722\\
	1.1	34.7920984264361\\
	1.11	35.4339655618001\\
	1.12	36.0758326971641\\
	1.13	36.7176998325281\\
	1.14	37.359566967892\\
	1.15	38.0014341032561\\
	1.16	38.64330123862\\
	1.17	39.285168373984\\
	1.18	39.927035509348\\
	1.19	40.5689026447119\\
	1.2	41.210769780076\\
	1.21	41.8526369154399\\
	1.22	42.4945040508039\\
	1.23	43.1363711861679\\
	1.24	43.7782383215319\\
	1.25	44.4201054568959\\
	1.26	45.0619725922599\\
	1.27	45.7038397276238\\
	1.28	46.3457068629878\\
	1.29	46.9875739983518\\
	1.3	47.6294411337158\\
	1.31	48.2713082690798\\
	1.32	48.9131754044437\\
	1.33	49.5550425398077\\
	1.34	50.1969096751717\\
	1.35	50.8387768105357\\
	1.36	51.4806439458997\\
	1.37	52.1225110812637\\
	1.38	52.7643782166276\\
	1.39	53.4062453519917\\
	1.4	54.0481124873556\\
	1.41	54.6899796227196\\
	1.42	55.3318467580836\\
	1.43	55.9737138934476\\
	1.44	56.6155810288115\\
	1.45	57.2574481641756\\
	1.46	57.8993152995395\\
	1.47	58.5411824349035\\
	1.48	59.1830495702675\\
	1.49	59.8249167056315\\
	1.5	60.4667838409954\\
};
\addlegendentry{All features}

\addplot [color=mycolor1, line width=1.5pt]
  table[row sep=crcr]{%
  	0   35.95\\
0.01	35.2771874516363\\
0.02	34.7410884396707\\
0.03	34.2049894277052\\
0.04	33.6688904157396\\
0.05	33.1327914037741\\
0.06	32.5966923918085\\
0.07	32.0605933798429\\
0.08	31.5244943678774\\
0.09	30.9883953559118\\
0.1	30.4522963439462\\
0.11	29.9161973319807\\
0.12	29.3800983200151\\
0.13	28.8439993080496\\
0.14	28.307900296084\\
0.15	27.7718012841185\\
0.16	27.2357022721529\\
0.17	26.6996032601873\\
0.18	26.1635042482218\\
0.19	25.6274052362562\\
0.2	25.0913062242907\\
0.21	24.5552072123251\\
0.22	24.0191082003596\\
0.23	23.483009188394\\
0.24	22.9469101764284\\
0.25	22.4108111644629\\
0.26	21.8747121524973\\
0.27	21.3386131405318\\
0.28	20.8025141285662\\
0.29	20.2664151166007\\
0.3	19.7303161046351\\
0.31	19.1942499277546\\
0.32	18.6601485963229\\
0.33	18.1260804053829\\
0.34	17.5939175866936\\
0.35	17.0641735167327\\
0.36	16.5375733980605\\
0.37	16.0152917032718\\
0.38	15.5014728274466\\
0.39	14.999665745946\\
0.4	14.5005089360886\\
0.41	14.0096068487029\\
0.42	13.5240578459479\\
0.43	13.0456553875269\\
0.44	12.5768773185822\\
0.45	12.1238461646895\\
0.46	11.6838179101985\\
0.47	11.2665374875982\\
0.48	10.8823847449563\\
0.49	10.5225113479121\\
0.5	10.1824945302908\\
0.51	9.86131644441429\\
0.52	9.56081784006331\\
0.53	9.27474498708026\\
0.54	8.99922402145588\\
0.55	8.73849171243123\\
0.56	8.49256345912327\\
0.57	8.26845900886361\\
0.58	8.0646973704811\\
0.59	7.8800459624553\\
0.6	7.7231241320441\\
0.61	7.59094025274086\\
0.62	7.47787335480725\\
0.63	7.39718940183999\\
0.64	7.34459495867374\\
0.65	7.31522090678417\\
0.66	7.31110310258069\\
0.67	7.33003656125073\\
0.68	7.37283055474761\\
0.69	7.44448756294775\\
0.7	7.53441656440045\\
0.71	7.64199655881771\\
0.72	7.76949251064721\\
0.73	7.91458987523117\\
0.74	8.07878560576328\\
0.75	8.26514108447302\\
0.76	8.46336336329723\\
0.77	8.68166573788238\\
0.78	8.92369141431218\\
0.79	9.17919632588738\\
0.8	9.45209091213396\\
0.81	9.73397077277929\\
0.82	10.0344767997044\\
0.83	10.3521817107677\\
0.84	10.6835518876048\\
0.85	11.03225164448\\
0.86	11.3904800654177\\
0.87	11.7641424151122\\
0.88	12.1596752112982\\
0.89	12.5693419717408\\
0.9	12.9959556578866\\
0.91	13.4347475298645\\
0.92	13.8862005362438\\
0.93	14.3465248576514\\
0.94	14.8217709685138\\
0.95	15.31060643509\\
0.96	15.8090411804712\\
0.97	16.3133782322428\\
0.98	16.8263501972552\\
0.99	17.3411468735094\\
1	17.8563651205652\\
1.01	18.3750226877278\\
1.02	18.8968598223163\\
1.03	19.4228037315591\\
1.04	19.9516371082607\\
1.05	20.4842908775311\\
1.06	21.018591251425\\
1.07	21.5528916253189\\
1.08	22.0871919992129\\
1.09	22.621505840644\\
1.1	23.1576048526096\\
1.11	23.6937038645751\\
1.12	24.2298028765406\\
1.13	24.7659018885062\\
1.14	25.3020009004718\\
1.15	25.8380999124373\\
1.16	26.3741989244029\\
1.17	26.9102979363685\\
1.18	27.446396948334\\
1.19	27.9824959602996\\
1.2	28.5185949722651\\
1.21	29.0546939842307\\
1.22	29.5907929961962\\
1.23	30.1268920081618\\
1.24	30.6629910201273\\
1.25	31.199090032093\\
1.26	31.7351890440585\\
1.27	32.271288056024\\
1.28	32.8073870679896\\
1.29	33.3434860799551\\
1.3	33.8795850919207\\
1.31	34.4156841038863\\
1.32	34.9517831158518\\
1.33	35.4878821278174\\
1.34	36.0239811397829\\
1.35	36.5600801517485\\
1.36	37.0961791637141\\
1.37	37.6322781756796\\
1.38	38.1683771876451\\
1.39	38.7044761996107\\
1.4	39.2405752115762\\
1.41	39.7766742235418\\
1.42	40.3127732355074\\
1.43	40.848872247473\\
1.44	41.3849712594385\\
1.45	41.9210702714041\\
1.46	42.4571692833696\\
1.47	42.9932682953352\\
1.48	43.5293673073007\\
1.49	44.0654663192663\\
1.5	44.6015653312318\\
};
\addlegendentry{Subset}

\end{axis}
\end{tikzpicture}%
	\vspace{-0.9cm}
	\caption{Mean estimation error $\bar{\varepsilon}$ for the validation of the Conventional10 setup for two models depending on the value of $\zeta$. The energy coefficients $e_i$ are  trained on the Conventional8 setup. First, all features of the FU model are influenced by the bit depth (red curve). Second, only a subset of features of the FU model (cf. $\phi_i$ in Table~\ref{tab:FEModel}) is influenced by the bit depth (blue curve).} 
	\label{fig:EnergyConversion}
\end{figure}

Since the estimation of bit streams with a different bit depth is insufficient, we extended the energy model of \eqref{eq:51} with new parameters (cf. Section~\ref{subsec:3B}). Therefore, the improvement of $\phi_i$, which indicates whether a feature is influenced by the bit depth, will be discussed first. In Figure~\ref{fig:EnergyConversion}, two assumptions are compared. First, all specific energy coefficients depend on the bit depth ($\phi_{i}=1,\forall i$; see red curve). Second, only a subset of features \mbox{(cf. $\phi_i$ in Table~\ref{tab:FEModel})} depends on the bit depth (blue curve). The x-axis is the value of $\zeta$ and the y-axis is the mean estimation error $\bar{\varepsilon}$. The graph shows that the lowest $\bar{\varepsilon}$ of the red curve is $10.61\%$ ($\zeta=0.56$) and of the blue curve is $7.31\%$ ($\zeta=0.66$). This significant difference of $\bar{\varepsilon}$ suggests that the energy difference can be modeled by scaling a subset of the energy coefficients.

\begin{table}[!t]
	\vspace{-0.5cm}
	\caption{Mean estimation error for the validation of the 10-bit setups. The specific energy coefficients are trained with the Conventional8 setup and the FU model. Furthermore, the value of $\zeta$ is either $0$ (c.f. Tab.~\ref{tab:FirstResults}) or $0.66$.}
	\centering
	\bgroup
	\def\arraystretch{1}
	\vspace{-0.2cm}
	\begin{tabular}{c | c | c }
		& \multicolumn{2}{c}{Training: Conventional8 } \\
		Validation Setup &  ~~$\zeta=0$~~   & $\zeta=0.66$  \\	                 
		\hline
		Conventional10    &  $35.95\%$   &   $7.31\%$ \\
		360D10            &  $38.29\%$  &   $6.52\%$ \\
		HDR10             &  $40.38\%$  &   $9.89\%$  
	\end{tabular}
	\label{tab:Optimized with Zeta}
	\egroup
\vspace{-0.6cm}
\end{table}

Table~\ref{tab:Optimized with Zeta} compares the estimation errors of the proposed extension ($\zeta=0.66$) with the results from Table~\ref{tab:FirstResults} under the assumption that the Conventional8 setup is used for training. We can see that all estimation errors are significantly reduced to values below $10\%$. These results show that with the proposed extended model, a conventional SDR training setup is sufficient for the estimation of unconventional video formats, if mean errors of $10\%$ are acceptable.

\vspace{-0.1 cm}
\section{Conclusion}
\vspace{-0.1cm}
\label{sec:conclusion}
This paper showed that our proposed bit stream feature-based model with additional features can be used to accurately estimate the decoding energy of various unconventional video formats like 360$^\circ$, fisheye, and HDR. Using a suitable training set, mean estimation errors below $5\%$ can be achieved. If the model is solely trained on conventional sequences, estimation errors below $10\%$ were observed. The extension of the energy model showed that the additional energy demand of bit streams with a higher bit depth can also be modeled accurately. 

In future work, the model shall be used to reduce the decoding energy using, e.g., decoding-energy-rate-distortion optimization. Further research regarding the decoding energy of different bit depth shall investigate a generally applicable value for $\zeta$. Furthermore, the model shall be extended to the upcoming VVC video compression standard.

\vspace{-0.1cm}
\bibliographystyle{IEEEtran}

\begin{thebibliography}{10}
\providecommand{\url}[1]{#1}
\csname url@samestyle\endcsname
\providecommand{\newblock}{\relax}
\providecommand{\bibinfo}[2]{#2}
\providecommand{\BIBentrySTDinterwordspacing}{\spaceskip=0pt\relax}
\providecommand{\BIBentryALTinterwordstretchfactor}{4}
\providecommand{\BIBentryALTinterwordspacing}{\spaceskip=\fontdimen2\font plus
\BIBentryALTinterwordstretchfactor\fontdimen3\font minus
  \fontdimen4\font\relax}
\providecommand{\BIBforeignlanguage}[2]{{%
\expandafter\ifx\csname l@#1\endcsname\relax
\else
\language=\csname l@#1\endcsname
\fi
#2}}
\providecommand{\BIBdecl}{\relax}
\BIBdecl

\bibitem{CSI2019}
\BIBentryALTinterwordspacing
{Cisco}. (2019, Feb.) Cisco visual networking index: Global mobile data traffic
  forecast update, 2017-2022. [Online]. Available:
  \url{https://www.cisco.com/c/en/us/solutions/collateral/service-provider/visual-networking-index-vni/white-paper-c11-738429.pdf}
\BIBentrySTDinterwordspacing

\bibitem{Sullivan2017}
G.~J. {Sullivan}, J.~{Boyce}, and T.~{Wiegand}, ``Requirements for future video
  coding,'' International Telecommunication Union ({ITU}) Telecommunication
  Standardization Sector SG16-TD155-A2/PLEN, Macao, China, {Document
  ToR-JVET-TD-PLEN-0155A2-A20171027-Reqs-VCEG}, Oct. 2017.

\bibitem{JVET-H1002}
A.~Segall, V.~Baroncini, J.~Boyce, J.~Chen, and T.~Suzuki, ``Joint call for
  proposals on video compression with capability beyond {HEVC},'' Joint Video
  Exploration Team of ITU-T SG 16 WP 3 and ISO/IEC JTC 1/SC 29/WG 11, AHG
  Report, {JVET-H1002-v6}, Oct. 2017.

\bibitem{HerglotzSpringerReichenbachEtAl2018}
C.~{Herglotz}, D.~{Springer}, M.~{Reichenbach}, B.~{Stabernack}, and A.~{Kaup},
  ``Modeling the energy consumption of the {HEVC} decoding process,''
  \emph{IEEE Transactions on Circuits and Systems for Video Technology},
  vol.~28, no.~1, pp. 217--229, Jan. 2018.

\bibitem{HerglotzHeindelKaup}
C.~{Herglotz}, A.~{Heindel}, and A.~{Kaup}, ``Decoding-energy-rate-distortion
  optimization for video coding,'' \emph{IEEE Transactions on Circuits and
  Systems for Video Technology}, vol.~29, no.~1, pp. 171--182, Jan. 2019.

\bibitem{Herglotz2019}
\BIBentryALTinterwordspacing
C.~{Herglotz}. (2019) {Decoding Energy Estimation Tool (denesto)}. {accessed
  2019-06-05}. [Online]. Available: \url{https://denesto.lms.tf.fau.de/}
\BIBentrySTDinterwordspacing

\bibitem{HerglotzWenDaiEtAl2016}
C.~{Herglotz}, Y.~{Wen}, B.~{Dai}, M.~{Kr\"anzler}, and A.~{Kaup}, ``A
  bitstream feature based model for video decoding energy estimation,'' in
  \emph{Proc. Picture Coding Symposium (PCS)}, Nuremberg, Germany, Dec. 2016.

\bibitem{CorreaCorreaPalominoEtAl2018}
D.~{Corr\^ea}, G.~{Correa}, D.~{Palomino}, and B.~{Zatt}, ``{OTED}: Encoding
  optimization technique targeting energy-efficient {HEVC} decoding,'' in
  \emph{Proc. IEEE Intl. Symposium on Circuits and Systems (ISCAS)}, Florence,
  Italy, May 2018.

\bibitem{MallikarachchiTalagalaH.EtAl2017}
T.~Mallikarachchi, D.~S. Talagala, H.~K. {Arachchi}, and A.~Fernando, ``A
  feature based complexity model for decoder complexity optimized {HEVC} video
  encoding,'' in \emph{Proc. IEEE Intl. Conference on Consumer Electronics
  (ICCE)}, Las Vegas, NV, USA, Jan. 2017.

\bibitem{HM16}
\BIBentryALTinterwordspacing
(2018) {HEVC test model reference software (HM-16.16)}. {Accessed 2018-10-15}.
  [Online]. Available:
  \url{https://hevc.hhi.fraunhofer.de/svn/svn_HEVCSoftware/tags/HM-16.16/}
\BIBentrySTDinterwordspacing

\bibitem{JCTVC-L1100}
F.~Bossen, ``Common test conditions and software reference configurations,''
  Joint Collaborative Team on Video Coding of ITU-T SG16 WP3 and ISO/IEC
  JTC1/SC29/WG11, {Document JCTVC-L1100}, Jan. 2013.

\bibitem{JVET-K1010}
F.~Bossen, J.~Boyce, K.~Suehring, X.~Li, and V.~Seregin, ``{JVET} common test
  conditions and software reference configurations for {SDR} video,'' Joint
  Video Exploration Team of ITU-T SG 16 WP 3 and ISO/IEC JTC 1/SC 29/WG 11,
  {Document JVET-K1010-v2}, Jul. 2018.

\bibitem{JVET-K1011}
A.~Segall, E.~François, and D.~Rusanovskyy, ``{JVET} common test conditions
  and evaluation procedures for {HDR/WCG} video,'' Joint Video Exploration Team
  of ITU-T SG 16 WP 3 and ISO/IEC JTC 1/SC 29/WG 11, Ljubljana, Slovenia,
  {Document JVET-K1011}, Jul. 2018.

\bibitem{Eichenseer2016}
A.~Eichenseer and A.~Kaup, ``A data set providing synthetic and real-world
  fisheye video sequences,'' in \emph{Proc. IEEE Intl. Conf. on Acoustics,
  Speech and Signal Processing (ICASSP)}, Shanghai, China, Mar. 2016.

\bibitem{JVET-K1012}
P.~Hanhart, J.~Boyce, and K.~Choi, ``{JVET common test conditions and
  evaluation procedures for 360$^\circ$ video},'' Joint Video Exploration Team
  of ITU-T SG 16 WP 3 and ISO/IEC JTC 1/SC 29/WG 11, Ljubljana, Slovenia,
  {Document JVET-K1012-v1}, Jul. 2018.

\bibitem{360Lib}
\BIBentryALTinterwordspacing
(2018) 360lib projection format conversion software. {[Accessed 2018-10-15]}.
  [Online]. Available: \url{https://jvet.hhi.fraunhofer.de/svn/svn_360Lib}
\BIBentrySTDinterwordspacing

\bibitem{Sze2014}
V.~Sze, M.~Budagavi, and G.~J. Sullivan, \emph{High Efficiency Video Coding
  ({HEVC})}.\hskip 1em plus 0.5em minus 0.4em\relax New York, NY, USA:
  Springer-Verlag, 2014.

\bibitem{Raspberry}
\BIBentryALTinterwordspacing
(2019) {Raspberry Pi 3 Model B+}. {Accessed 2019-03-19}. [Online]. Available:
  \url{www.raspberrypi.org/products/raspberry-pi-3-model-b-plus/}
\BIBentrySTDinterwordspacing

\bibitem{Raspbian2}
\BIBentryALTinterwordspacing
(2018) {Raspbian}. {Accessed 2018-10-15}. [Online]. Available:
  \url{https://www.raspberrypi.org/downloads/raspbian/}
\BIBentrySTDinterwordspacing

\bibitem{FFmpeg}
\BIBentryALTinterwordspacing
(2019) {Fast Forwards MPEG (FFmpeg)}. Accessed 2018-11-14. [Online]. Available:
  \url{http://ffmpeg.org/}
\BIBentrySTDinterwordspacing

\bibitem{ColemanLi1996}
T.~F. {Coleman} and Y.~{Li}, ``An interior trust region approach for nonlinear
  minimization subject to bounds,'' \emph{SIAM Journal on optimization},
  vol.~6, no.~2, pp. 418--445, 1996.

\end{thebibliography}

\end{document}